\documentclass[twocolumn]{aastex61}

\newcommand\aastex{AAS\TeX}

\newcommand{\gps}{\ensuremath{g_{\rm P1}}}
\newcommand{\rps}{\ensuremath{r_{\rm P1}}}
\newcommand{\ips}{\ensuremath{i_{\rm P1}}}
\newcommand{\zps}{\ensuremath{z_{\rm P1}}}
\newcommand{\yps}{\ensuremath{y_{\rm P1}}}

\newcommand{\grizy}{\gps\rps\ips\zps\yps}

\submitjournal{ApJ}

\shorttitle{\aastex\  The Pan-STARRs1 Medium-deep Survey}
\shortauthors{Jian et al.}

\begin{document}

\title{The Pan-STARRS1 Medium-deep Survey: Star Formation Quenching in Group and Cluster Environments}

\correspondingauthor{Hung-Yu Jian}
\email{hyjian@asiaa.sinica.edu.tw}

\author{Hung-Yu Jian}
\affiliation{Institute of Astronomy \& Astrophysics, Academia Sinica, 106, Taipei, Taiwan, R.O.C.}
\affiliation{Department of Physics, National Taiwan University, 106, Taipei, Taiwan, R.O.C.}

\author{Lihwai Lin}
\affiliation{Institute of Astronomy \& Astrophysics, Academia Sinica, 106, Taipei, Taiwan, R.O.C.}

\author{Kai-Yang Lin}
\affiliation{Institute of Astronomy \& Astrophysics, Academia Sinica, 106, Taipei, Taiwan, R.O.C.}

\author{Sebastien Foucaud}
\affiliation{Department of Earth Sciences, National Taiwan Normal University, N.88, Tingzhou Road, Sec. 4, Taipei 11677, Taiwan, R.O.C.}
\affiliation{Shanghai Jiao Tong University, Shanghai, China}

\author{Chin-Wei Chen}
\affiliation{Institute of Astronomy \& Astrophysics, Academia Sinica, 106, Taipei, Taiwan, R.O.C.}

\author{Tzihong Chiueh}
\affiliation{Department of Physics, National Taiwan University, 106, Taipei, Taiwan, R.O.C.}
\affiliation{Center for Theoretical Sciences, National Taiwan University, 106, Taipei, Taiwan, R.O.C. }
\affiliation{LeCosPa, National Taiwan University, 106, Taipei, Taiwan, R.O.C.}

\author{R. G. Bower}
\affiliation{Institute for Computational Cosmology, Department of Physics, Durham University, South Road, Durham DH1 3LE, UK}

\author{Shaun Cole}
\affiliation{Institute for Computational Cosmology, Department of Physics, Durham University, South Road, Durham DH1 3LE, UK}

\author{Wen-Ping Chen}
\affiliation{Graduate Institute of Astronomy, National Central University, Chung-Li 32054, Taiwan, R.O.C.}

\author{W. S. Burgett}
\affiliation{Institute for Astronomy, University of Hawaii at Manoa, Honolulu, HI 96822, USA}

\author{P. W. Draper}
\affiliation{Institute for Computational Cosmology, Department of Physics, Durham University, South Road, Durham DH1 3LE, UK}

\author{H. Flewelling}
\affiliation{Institute for Astronomy, University of Hawaii at Manoa, Honolulu, HI 96822, USA}

\author{M. E. Huber}
\affiliation{Institute for Astronomy, University of Hawaii at Manoa, Honolulu, HI 96822, USA}

\author{N. Kaiser}
\affiliation{Institute for Astronomy, University of Hawaii at Manoa, Honolulu, HI 96822, USA}

\author{R.-P. Kudritzki}
\affiliation{Institute for Astronomy, University of Hawaii at Manoa, Honolulu, HI 96822, USA}

\author{E. A. Magnier}
\affiliation{Institute for Astronomy, University of Hawaii at Manoa, Honolulu, HI 96822, USA}

\author{N. Metcalfe}
\affiliation{Institute for Computational Cosmology, Department of Physics, Durham University, South Road, Durham DH1 3LE, UK}

\author{R. J. Wainscoat}
\affiliation{Institute for Astronomy, University of Hawaii at Manoa, Honolulu, HI 96822, USA}

\author{C. Waters}
\affiliation{Institute for Astronomy, University of Hawaii at Manoa, Honolulu, HI 96822, USA}



\begin{abstract}

We make use of a catalog of 1600 Pan-STARRS1 groups produced by the probability friends-of-friends algorithm to explore how the galaxy properties, i.e. the specific star formation rate (SSFR) and quiescent fraction, depend on stellar mass and group-centric radius. The work is the extension of \cite{lin14}. In this work, powered by a stacking technique plus a background subtraction for contamination removal, a finer correction and more precise results are obtained than in our previous work. We find that while the quiescent fraction increases with decreasing group-centric radius the median SSFRs of star-forming galaxies in groups at fixed stellar mass drop slightly from the field toward the group center.  This suggests that the major quenching process in groups is likely a fast mechanism. On the other hand, a reduction in SSFRs by $\sim$0.2 dex is seen inside clusters as opposed to the field galaxies. If the reduction is attributed to the slow quenching effect, the slow quenching process acts dominantly in clusters. In addition, we also examine the density$-$color relation, where the density is defined by using a sixth-nearest neighbor approach. Comparing the quiescent fractions contributed from the density and radial effect, we find that the density effect dominates over the massive group or cluster galaxies, and the radial effect becomes more effective in less massive galaxies. The results support mergers and/or starvation as the main quenching mechanisms in the group environment, while harassment and/or starvation dominate in clusters. 

\end{abstract}





\section{Introduction}
Galaxy properties exhibit strong correlations with their stellar mass ($M_{*}$) and hosting environments \citep{dre80,coo07,ger07,pen10,muz12,wet12}. Galaxies in denser environments tend to be older, redder, and less strongly star-forming. However, more massive galaxies also have a tendency to be older, redder, and have less strongly star-forming active star formation \citep{bal06,pen10}. The debate about the extent to which the properties of galaxies are determined by external processes (environmental quenching, associated with a galaxy being accreted by a cluster or group and evolving as a satellite) or internal processes (assumed to be tied to galaxy mass) is ongoing. To understand the relative influence of stellar mass and environment on galaxy evolution, it is necessary to isolate the processes responsible for the observed relation of star formation rate (SFR) to density. If the main cause is primarily internal processes, then it is the difference in the galaxy stellar mass distributions in different environments that leads to the SFR$-$density relation. In contrast, the environmental process alters the properties of galaxies of a given stellar mass. Therefore, by probing the SFR$-M_{*}$ relation, we can obtain insights into these two physical processes.

In our previous study, \cite{lin14}, we confirmed that the group environment strongly affects the fraction of quiescent galaxies at fixed stellar mass, but no environmental dependence was found for the star-forming sequence, in good agreement with previous works \citep{bal04,vul10,koy13}. The result thus supports a fast environment-quenching scenario and favors galaxy mergers in groups to be the primary quenching mechanism. On the other hand, for the cluster sample, we found a global reduction of 17$\%$ in the specific star formation rate (SSFR) of the star-forming sequence compared to its field counterpart, consistent with many cluster studies \citep{vul10,hai13,alb14}. The excess of the quiescent fraction in clusters compared to that in the field galaxies reveals a positive correlation with the stellar mass similar to the trend of the mass-dependent environment quenching in the strangulation model introduced by \cite{font08}. It is thus concluded that the quenching mechanism of strangulation appears to fit better the observed trend better.

It is known that an evolutionary sequence of star formation is likely related to radial gradients through the dynamical friction, which causes the correlation between satellite radius and infall time \citep{gao04}. Hence galaxy properties as a function of the radius can provide us with hints of the relative importance of different quenching processes that halt star formation in a dense environment, and further help our understanding of galaxy formation and evolution. We hence extend our previous results to include the discussion of galaxy properties in terms of the projected group-centric radius $r_p$ and aim to understand at which radius the possible quenching mechanisms acting on galaxies effectively alter galaxy properties. This work is based on the same group catalog of two PS1 medium-deep fields, MD04 and MD07, produced by the probability friends-of-friends (or PFOF) algorithm in \cite{lin14}. By taking advantage of the stacking technique plus background subtraction and removal of the field contamination, we obtain a finer correction for our stacked samples and thus more robust results than in our previous work. Dependencies of galaxy properties can therefore be probed in detail. In this work, we present our results on the SSFR, the quiescent fraction, and the environmental quenching efficiency as functions of $r_p$ in different stellar mass bins to uncover the underlying SSFR, $M_{*}$, and $r_p$ distribution of the group or/and cluster galaxies and to understand the possible mechanisms causing environmental quenching in groups or/and clusters. In addition, using the same PS1 MD galaxy catalogs, we also explore the color$-$density relation where the density is estimated following the $n$th-nearest-neighbor approach \citep[e.g.][]{coo07}, and direct comparisons are made between two different approaches to understand the separate contributions from the density and radial effect, respectively.

Our paper is organized as follows. In Section 2, we briefly describe the data. The analysis method is then illustrated in Section 3. In Section 4, we present the main results. Section 5 discusses the important implications of our results for understanding the evolution of galaxies. We adopt the following cosmology throughout this paper: $H_0$ = 100 $h~\textrm{km}~s^{-1}~\textrm{Mpc}^{-1}$, $\Omega_\textrm{m}$= 0.3, and $\Omega_{\Lambda}$ = 0.7. We adopt the Hubble constant $h$ = 0.7 when calculating rest-frame magnitudes. We use a Salpeter initial mass function (IMF) when deriving stellar masses and SFRs. All magnitudes are given in the AB system.

\section{Data}
The galaxy and group sample analyzed in this work are the same as was used in \cite{lin14} with an $i$-band selection cut, $i$ $<$ 24.0. We briefly describe the data here. The detailed description of the galaxy sample can be found in S. Foucaud et al. (2017, in preparation) and in \cite{lin14}, and for that of the group/cluster sample is described in \cite{jian14}.

\subsection{Galaxy Sample}

Our galaxy sample is based on images and photometry from the Pan-STARRS1 \citep[PS1;][]{kai10} Medium-Deep (MD) Field survey with $\grizy$ filters \citep{stu10,ton12} plus the Canada$-$France$-$Hawaii$-$Telescope (CFHT) MEGACAM $u^*$ band from archival data from a program lead by Eugene Magnier as part of the PS1 efforts to produce deep stacks in six bands and associated catalogs. Data taken in two PS1 MD fields, MD04 and MD07, are adopted for use (S. Foucaud et al. 2017, in preparation).

With the six optical bands including the PS1 $\grizy$ band and CFHT $u^*$ photometry, photometric redshifts (hereafter photo-$z$) are computed by fitting with the publicly available EAZY code \citep{bra08}, adopting the template set called ``CFHTLS-SED'' from a public photo-$z$ software ``LePhare'' \citep{arn99,ilb06}. Additionally, a prior is applied on the redshift distribution for any given range of $i$-band magnitude using a mock galaxy catalog that assumes the semi-analytical galaxy formation model of \cite{guo10}. The spectroscopic redshift samples of zCOSMOS \citep{lil07} and DEEP2 \citep{new13} are used to calibrate the zero-points and to characterize the photo-$z$ performances in MD04 and MD07, respectively. The photo-$z$ accuracy and the outlier rate are 0.047 and 4$\%$ at $\ips$ $<$ 22.5 in MD04, and 0.051 and 7$\%$ at $\rps$ $<$ 24.1 in MD07 \citep{lin14}. Moreover, stellar masses are derived by using the spectral energy distribution (SED) fitting code ``FAST'' \citep{kri09} assuming the models of \cite{bru03}. The stellar mass completeness of our sample is found to be $log_{10}(M_{*}/M_{\odot})$ = 9.4 (9.0), 10.1 (9.7), and 10.5 (10.1) for red (blue) galaxies at $z$ $\sim$ 0.2, $z$ $\sim$ 0.5, and $z$ $\sim$ 0.8, respectively \citep{lin14}.

Finally, the SFR is derived by adopting the approach described in \cite{mos12}. They parameterize the SFR with a combination of observed quantities, including rest-frame optical $B$ magnitudes $M_B$, observed [O$_{\rm{II}}$] luminosity, and/or rest-frame color. The parameterized SFR is then calibrated against the SED-fit SFR from UV/optical bands in the AEGIS survey in the redshift range of 0.7 $<$ $z$ $<$ 1.4. By incorporating a dimming factor of $Q$ = 1.3 magnitudes per unit redshift for the measured values $M_B$ to correct for dust extinction, Mostek and collaborators  found that the derived $z$ $\sim$ 1 SFRs agree well with the local L[O$_{\rm{II}}$]$-$M$_B$ SFR calibration commonly used in the literature \citep{kew04}.  Following the work in \cite{lin14}, we use the fitting formula employing the rest-frame optical $M_B$, $(U-B)$, and a second order $(U-B)$ color as fit parameters. The fitting coefficients can be found in Table 3 in \citet{mos12}. \cite{mos12} reported that even though the [O$_{\rm{II}}$] flux of red galaxies may have contributions from AGN activity, this effect is much smaller for star-forming galaxies. Since our main purpose is to explore the star formation rate for the star-forming main sequence and the quiescent fraction in different environments, we do not rely on the SFR measurements for the quiescent galaxies at all, and hence the AGN effect can be ignored in this work.

\citet{mos12} also reported that the galaxy color depends on the SFR uncertainties. The adopted fitting formula in this work is found to have a scatter of residual $\sim$ 0.19 for the star-forming galaxies and $\sim$ 0.45 for the quiescent galaxies. That is, the method may not yield precise SFR measurements for star-forming and quiescent galaxies.  To estimate the uncertainty in our analysis that is introduced by this method, we have performed Monte Carlo simulations. Given a median SSFR for the quiescent and star-forming galaxies and a quiescent fraction, we construct a distribution of SSFR with double-Gaussian peaks to mimic distinct star-forming and quiescent populations. We then randomly perturb the SSFRs with a scatter 0.45 and 0.19 for the quiescent and star-forming population, respectively, and repeat the same procedure for ten thousand times to obtain the bias and deviation for the median SSFR of quiescent and star-forming galaxies and the quiescent fraction relative to our original inputs. Given different inputs based on our results in four different mass ranges, including the median SSFR for the quiescent and star-forming galaxies and the corresponding quiescent fraction, we find that the bias and scatter for the median SSFR of star-forming galaxies caused by the method are -0.04 dex and 0.006 for high-mass galaxies and -0.02 dex and 0.003 for low-mass galaxies, which is much less lower than the errors from the jackknife resampling. Additionally, the bias for the quiescent fraction can be as large as -8$\%$ (-0.07) for high-mass galaxies and -4$\%$ (-0.02) for low-mass galaxies, and the corresponding scatter is an order of magnitude smaller than the bias, $\sim$2-3 $\times$10$^{-3}$. The introduced bias and scatter appear to be significantly smaller than the errors from the jackknife resampling. We thus neglect to include the uncertainties in the error bars.

\subsection{Group/Cluster Sample}
The group/cluster sample is constructed using the group identification method called the probability friend-of-friend group finder \citep[PFOF;][]{liu08,jian14}. On the basis of the conventional FOF method \citep{huc82}, PFOF additionally takes into account the probability distribution function of the photometric redshifts of each galaxy and computes the linking probability of a given pair of galaxies to quantify their association in the line-of-sight direction. Given a linking probability threshold along with the linking lengths, PFOF then identifies groups and clusters. With a training set containing known spectroscopically identified groups and clusters in the same field, we can optimize to obtain an optimal product.

In this work, we make use of an updated version of the PFOF-generated group samples in MD04 and MD07 \citep{lin14}, which are based on a set of linking lengths and threshold trained by using the spectroscopically identified group sample from DEEP2 \citep{ger12} in the PS1 MD07 field. The PFOF group samples are divided into two subsets, one with a richness of between 10 and 25 as the ``group'' sample and the other with a richness $> 25$ as the ``cluster'' sample. The richness cut roughly corresponds to a mass of $10^{13.2} < M_{\textrm{halo}} < 10^{13.8} M_{\odot}$ at $z$ $\sim$ 0.4 and $10^{13.4} < M_{\textrm{halo}} < 10^{14.0} M_{\odot}$ at $z$ $\sim$ 0.8, respectively. The combined catalog consists of 610 groups and 76 clusters at low redshift, and 875 groups and 61 clusters at high redshift.

\begin{figure}
\includegraphics[scale=0.7]{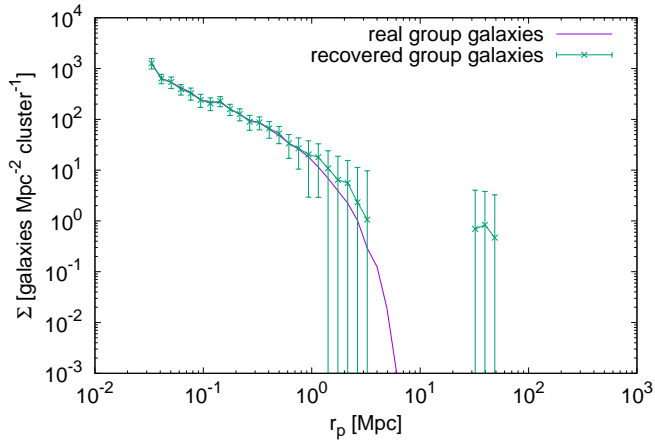}
\caption{Galaxy surface density $\Sigma$ is plotted as a function of projected radius $r_p$ for the sample using real group galaxies (purple) and recovered group galaxies (green). The recovered surface density shows good consistency with the real surface density.}
\label{f1}
\end{figure}

\begin{figure*}
\includegraphics[scale=1.4]{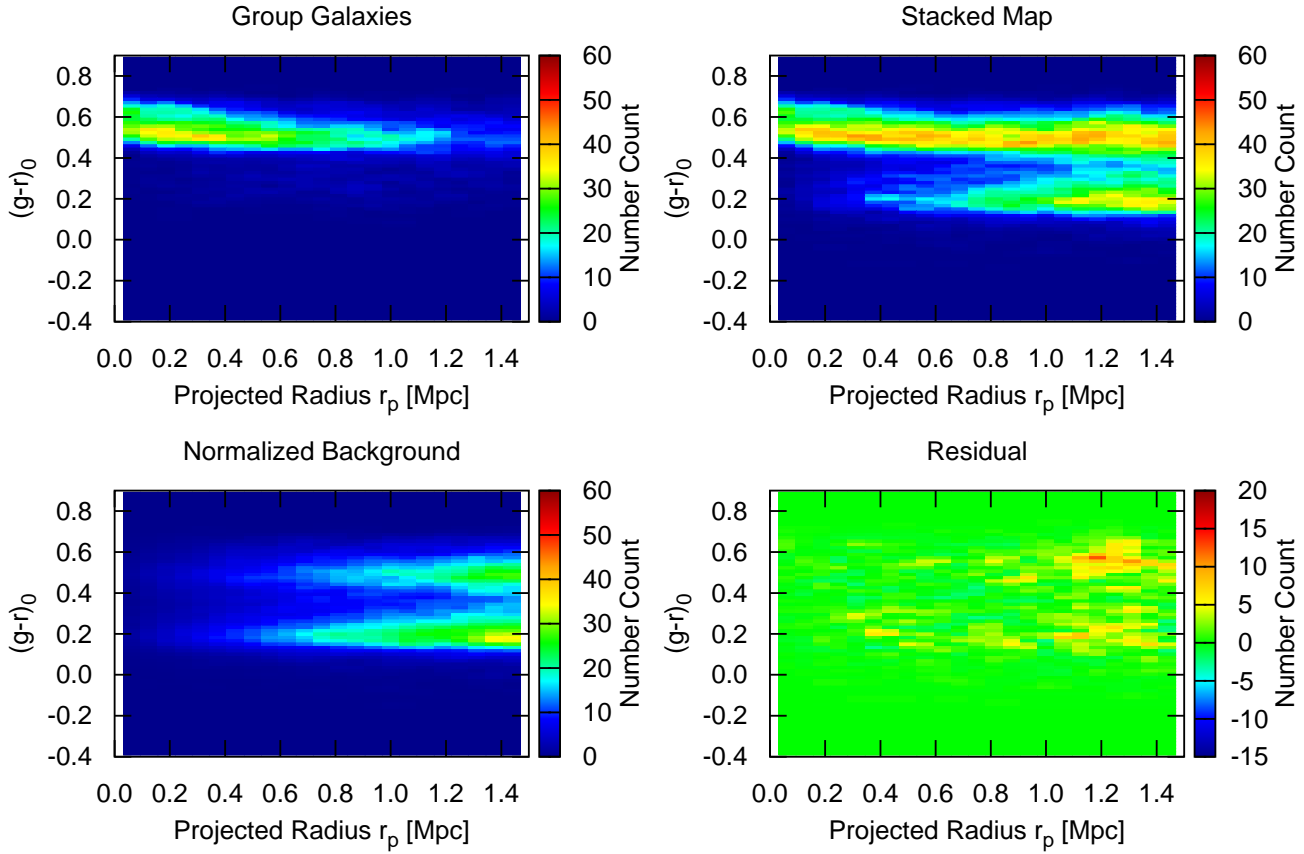}
\caption{Distribution of rest-frame $g$ minus $r$ color $(g-r)_0$ as a function of the projected radius $r_p$ shown as a color-coded number count for the sample using real group galaxies (upper-left panel), stacked galaxies (upper-right panel), normalized background galaxies (bottom-left panel), and the residual (bottom-right panel), where the residual is defined as the real group galaxies minus the recovered group galaxies.The residual has a mean 0.68 and a scatter 3.97, roughly $10\%$ of the maximum number count $\sim$ 45 in the real galaxy sample.}
\label{f2}
\end{figure*}

\begin{figure}
\includegraphics[scale=0.65]{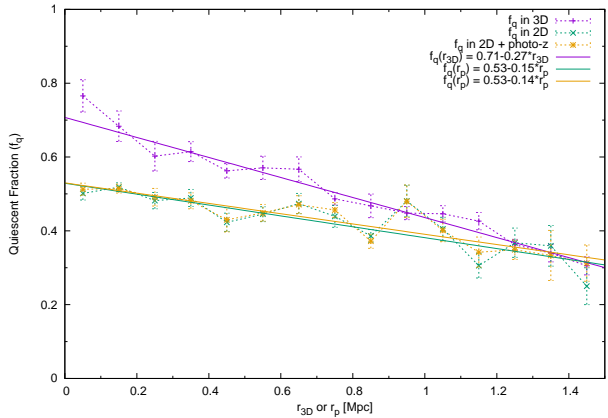}
\caption{Simulated quiescent fraction in 3D radius shown as a purple dashed line, while the green and gold dashed lines show the quiescent fractions in projection radius with spectral-$z$ and photo-$z$, respectively. Additionally, three solid lines represent the best-fitting results for the three different cases stated above. The projected $f_q$ for photo-$z$ galaxies agrees very well with that for spectral-$z$ galaxies, suggesting that the photo-$z$ uncertainty has little effect on the resulting $f_q$. If the background subtraction' can properly remove background galaxies, the projected $f_q$ in groups or in clusters can be well recovered.}
\label{f3}
\end{figure}

\begin{figure*}
\includegraphics[scale=1.4]{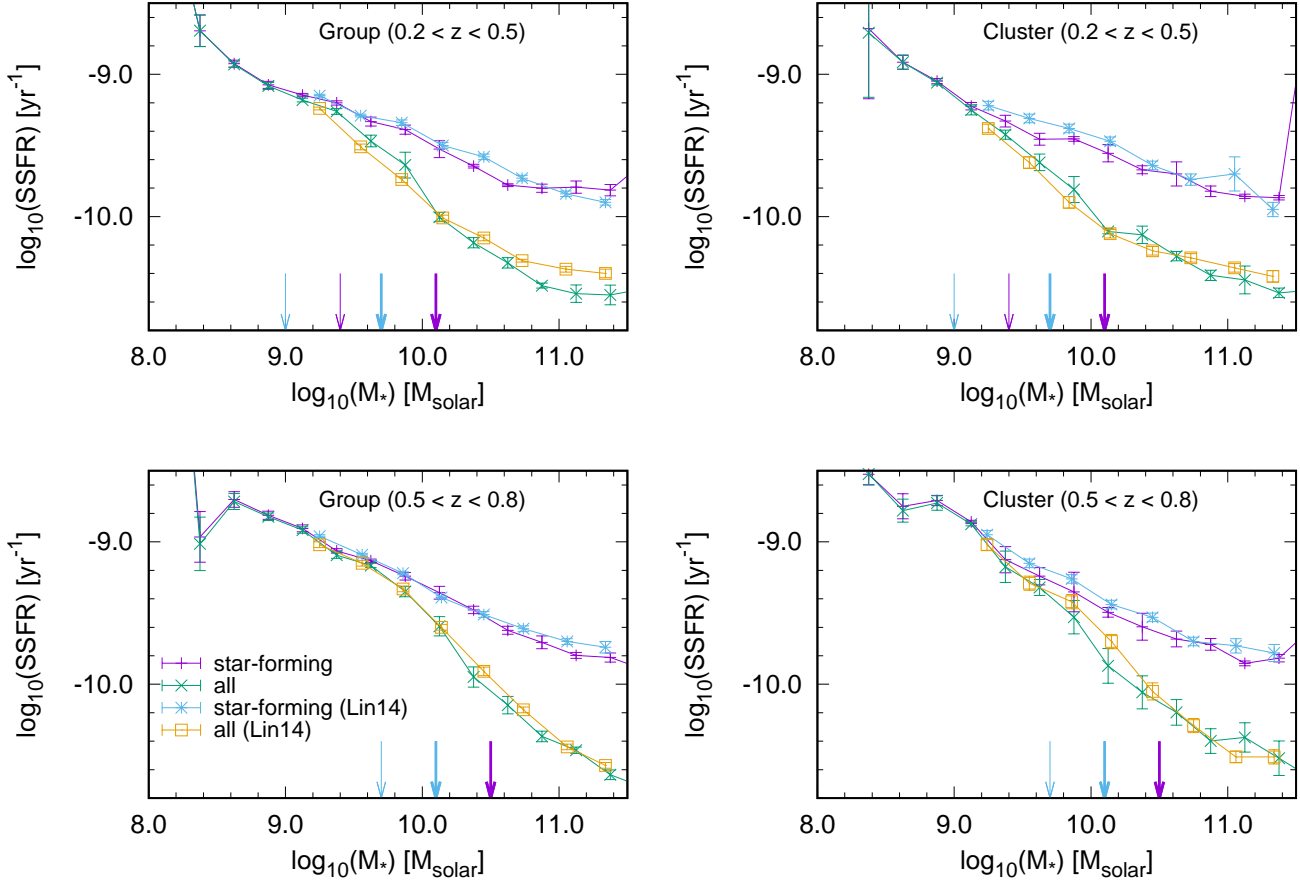}
\caption{Consistency check on the SSFR as a function of stellar mass between the approach using the PFOF group catalog \citep{lin14} and that using background subtraction. Blue stars and yellow squares with errors are the median SSFRs for the star-forming and for all galaxies from \cite{lin14}, respectively, while the purple (star-forming) and blue lines (all) are results from this work. The thick (thin) vertical arrows represent the mass completeness limits for galaxies with the reddest colors in the star-forming (blue) and quiescent (purple) populations at the upper- (lower-) redshift limits of each panel. It can be seen that the two independent methods show good consistency. }
\label{f4}
\end{figure*}

\begin{figure}
\includegraphics[scale=1.]{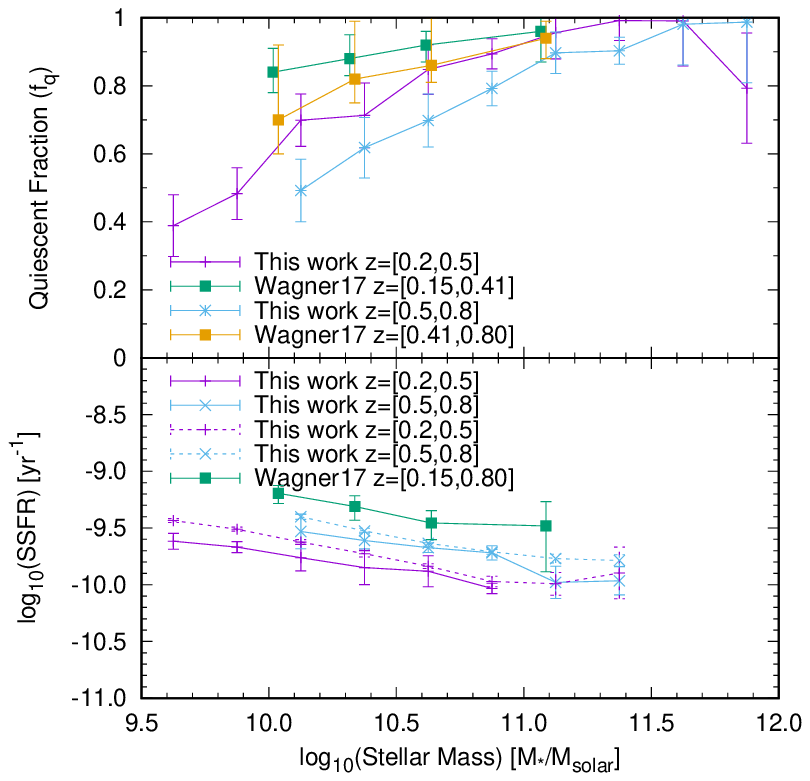}
\caption{ Top: the quiescent fraction $f_q$ as a function of the stellar mass $M_*$. The purple pluses and blue stars denote the $f_q$ in redshift range of 0.2 $< z <$ 0.5 and 0.5 $< z <$ 0.8, respectively. For comparisons, the $f_q$s from \cite{wag17} in  0.15 $<  z <$ 0.41 (green squares) and 0.41 $<  z <$ 0.80 (yellow squares) are included. Bottom: The SSFR of star-forming cluster (solid lines) and field (dashed lines) galaxies as a function of the stellar mass $M_*$ in 0.2 $< z <$ 0.5 (purple pluses) and 0.5 $< z <$ 0.8 (blue stars). The green squares show the SSFRs of star-forming cluster galaxies in 0.15 $< z <$ 0.8 from \cite{wag17}.}
\label{f5}
\end{figure}

\section{Method and Tests}

\subsection{Background Subtraction} \label{bs}
In this work, the method we adopt to correct for contamination in the stacking data is called ``background subtraction''. The background is decontaminated by considering a mean local background around each cluster in an annulus at a projected radius $r_1 < r_p < r_2$, where $r_1$ and $r_2$ are the inner and outer radii, respectively, and the contaminated galaxy properties at the center can then be removed by subtracting the same galaxy properties of the background. This method has been widely used to compute the luminosity function, e.g. \cite{bar07}, and has been demonstrated numerically to be able to accurately recover the underlying luminosity function selected in three dimensions \citep{val01}. We adopt the method to explore galaxy properties for photometric redshift samples, similar to the approach used in \cite{loh08} and \cite{li09}. We stack galaxies around group centers within a redshift slice for which the redshift difference between galaxy and group has to be smaller than the photo-$z$ accuracy, i.e. $|z - z_{\textrm{grp}}| \leq \sigma_{\Delta z/(1+z_s)}$, where $z$, $z_{\textrm{grp}}$, and $\sigma_{\Delta z/(1+z_s)}$ are galaxy redshift, group redshift, and photo-$z$ accuracy, respectively. The position of the brightest cluster galaxy (BCG) is selected as the center. From the center, the inner circle within the projection radius $r_p$ $\leq$ 1.5 $\textrm{Mpc}$ is considered as the stacked galaxy sample, including group and field galaxies, while the outer annulus between $r_1$ (8.0 $\textrm{Mpc}$) and $r_2$ (9.5 $\textrm{Mpc}$) is as the field galaxy (or background) sample. The recovered sample is then the difference between the stacked galaxy and the field sample normalized to have the same area as the stacked galaxy sample. In addition, we also construct a background sample selected from a random position within the same redshift width as the stacked group galaxy sample, and we find that the difference between the corrected sample using the annulus background and that using the random background is negligible, and the results show good consistency. 

\subsection{Stacking Tests Using a Mock Catalog} \label{mt}
To know the performance of the ``background subtraction'' method, we make use of a mock galaxy catalog, based on a semi-analytic galaxy formation model \citep{lag12}, constructed for the Pan-STARRS1 Medium Deep Surveys \citep[][also see Jian et al. 2014]{mer13}, with simulated photo-$z$ accuracy $\sigma_{\Delta z/(1+z_s)}$ $\sim$ 0.05 to test the method. We select clusters in the redshift range between $0.2 < z < 0.5$ with the host halo mass $M_h/M_{\odot} > 10^{14}$, and set the flux limit $i < 24.0$. We find 34 clusters and 7952 cluster galaxies from the mock catalog. For the first test, we stack photo-$z$ galaxies around these 34 clusters, subtract the background, and then compute the composite surface density as a function of $r_p$. We then compare the composite surface density from real group galaxies and from the recovered sample. The result is shown in Figure~\ref{f1}. In this case, the background sample is selected from random positions. The recovered surface density agrees well with the real surface density within 1 \textrm{Mpc} but at large radii beyond 1 \textrm{Mpc}, it exhibits a small excess with large error bars. 


The other test shown in Figure~\ref{f2} displays the rest-frame $g$ minus $r$ color, $(g-r)_0$, as a function of the projected group-centric radius $r_p$ as a two-dimensional color-coded number count map for the real group galaxies (upper-left panel), stacked galaxies (upper-right panel), normalized field galaxies (bottom-left panel), and residual (bottom-right panel), defined as the difference between real and recovered galaxy sample. The stacked galaxy sample is distinct from the real galaxy sample. However, after we subtract the normalized background sample from the stacked galaxy sample, the recovered sample shows similar structure to the real galaxy sample. The mean of the residual is 0.68 and the deviation is 3.97 in units of galaxies per group or cluster, roughly $10\%$ of the maximum number count $\sim$ 45 in the real galaxy sample. The above two tests demonstrate that the background subtraction method can successfully recover the underlying group galaxy properties.

Additionally, we also estimate the impact of the projection and photo-$z$s effects on the quiescent fraction. We first simulate group galaxies with a linearly decreasing quiescent fraction with increasing radius in 3D space. We then compute the quiescent fraction as a function of the projected radius to understand the projection effect. We also perturb each galaxy redshift to simulate the photo-$z$ effect and estimate the quiescent fraction as a function of the projected radius to mimic the case we will explore in this study. In Figure~\ref{f3}, we show a test result as an example, with the simulated quiescent fraction from 0.7 at the center to 0.3 at the boundary, 1.5 Mpc. Purple, blue, and yellow colors denote the cases using the 3D radius ($r_{3\textrm{D}}$), the projected radius, and the projected radius with a photo-$z$ inaccuracy $\sigma_{\Delta z/(1+z)} = 0.05$, respectively. Note that the 3D and projected radius are different, and the slope of the quiescent fractions is also different between the 3D and projected case. The solid lines represent the fits, while the dashed lines with the error bars show the data and statistical errors estimated from the deviation in 128 realizations. The quiescent fraction due to the projection and photo-$z$ effect clearly agree very well. The photo-$z$ effect appears to not significantly affect the quiescent fraction of the estimation. That is, if the background subtraction can properly recover the group members, it is expected that the resulting quiescent fraction retrieves the original 2D case. 




\subsection{Consistency Checks}
In our previous study, \cite{lin14}, we accounted for the contamination and incompleteness effects introduced in the PFOF group identification by estimating the rate of recovering true group memberships as well as the contamination rate from the field galaxies using galaxies that were cross-matched with spectroscopically identified group members. That is, the contamination removal method in our previous work is distinct from the current method. To understand the consistency between the two approaches, we compute the SSFR and the quiescent fraction as a function of stellar mass without the dimming factor correction via background subtraction and compare them with those in \cite{lin14}. The results are shown in Figure~\ref{f4}. Good consistency between the two approaches can be seen, demonstrating that our results are reliable. 

In addition, the SFR-$M_*$ relation of star-forming field galaxies in our sample are fit with a linear formula, i.e. log$_{10}$(SFR/$M_{\odot}$ yr$^{-1}$) = $\alpha$ $\times$ log$_{10}$ ($M_*$/$M_{\odot}$) + $\beta$, where $\alpha$ and $\beta$ are fitting parameters for the slope and amplitude, respectively. We find that in our sample, the best-fit parameters are $\alpha$ = 0.63 $\pm$ 0.02 and $\beta$ = -5.77 $\pm$ 0.25 in the redshift range of 0.2 $< z <$ 0.5, and $\alpha$ = 0.62 $\pm$ 0.03 and $\beta$ = -5.59 $\pm$ 0.29 in the range of 0.5 $< z <$ 0.8. For comparisons, the best-fit parameters $\alpha$ and $\beta$ are 0.65 and -6.10 at $z$ = 0.42, and 0.62 and -5.62 at $z$ = 0.63 using Equation (1)$-$(3) from \cite{whi12} with a Chabrier IMF, where $z$ = 0.42 and 0.63 is the median redshift of our sample in the range of 0.2 $< z <$ 0.5 and 0.5 $< z <$ 0.8, respectively. In addition, the best-fit $\alpha$ and $\beta$ from \cite{noe07} are 0.67 $\pm$ 0.08 and -6.19 $\pm$ 0.78 in 0.2 $< z <$ 0.7. The amplitudes from these two works are corrected for the difference in the IMF, i.e. 1.8 $M_{*,\textrm{C}}$ = $M_{*,\textrm{S}}$, where C and S denote the Chabrier and Salpeter IMF, respectively. Our best-fit parameters agree well with those from \cite{whi12} and \cite{noe07}.

Moreover, in Figure~\ref{f5}, the quiescent fraction (top) of cluster galaxies and SSFR (bottom) of star-forming cluster galaxies are displayed as a function of stellar mass and compared to those from \cite{wag17}. In the top panel, the purple pluses and blue stars are the quiescent fractions of cluster galaxies from this work in 0.2 $< z <$ 0.5 and 0.5 $< z <$ 0.8, respectively, while the green squares and yellow squares are those in the range of 0.15 $< z <$ 0.41 and 0.41 $< z <$ 0.80 from \cite{wag17}. The quiescent fractions in the two different redshift ranges from this work and from \cite{wag17} are both agree well. In the bottom panel, the purple pluses and blue crosses are the SSFRs of star-forming cluster (solid lines) and field (dashed lines) galaxies in the range of 0.2 $< z <$ 0.5 and 0.5 $< z <$ 0.8, respectively. The green squares are the SSFRs of the star-forming cluster galaxies from \cite{wag17} after correcting the mass difference. In Figure~\ref{f5}, our values of the quiescent fraction are uniformly lower than those from \cite{wag17} and the SSFRs of star-forming cluster galaxies from this study are also lower than those from \cite{wag17} with a systematics of $\sim$ 0.3 or 0.4 dex. The discrepancy is likely due to the different ways of separating star-forming and quiescent populations.  We adopt constant SSFRs,  while \cite{wag17} used the strength of the 4000 $\rm{\AA}$ break $D_n$(4000) to divide the two populations. 


\begin{figure*}
\includegraphics[scale=1.4]{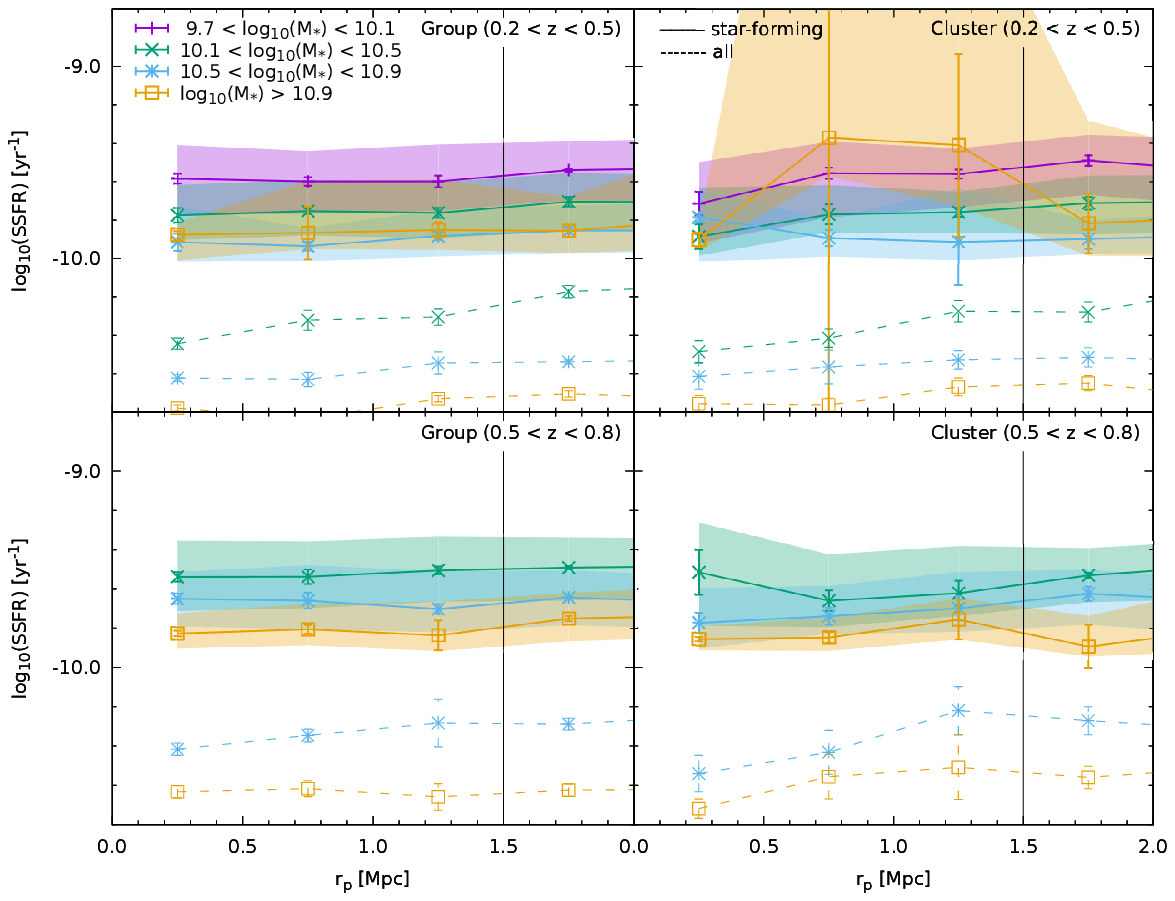}
\caption{Median specific star formation rate (SSFR) as a function of the group-centric radius $r_p$ in four different stellar mass ranges. The solid lines denote the star-forming population, while the dashed lines indicate all galaxies. The error bars are jackknife errors from eight subsamples. In addition, the black vertical lines indicate the boundary between groups (or clusters) and field. Owing to different mass completeness limits for the quiescent and star-forming population as well as for low and high redshift, at low-$z$ there are four bins for star-forming galaxies and three bins for all galaxies, while at high-$z$ there are three bins for star-forming galaxies and two bins for all galaxies. The shaded regions denote the SSFR of star-forming group or cluster galaxies between the 30$^{th}$ and 70$^{th}$ percentile. It is seen that the median SSFR of star-forming galaxies is less dependent on $r_p$ at all stellar masses in groups, while there is a more apparent decrease in SSFR toward the center in clusters.} 
\label{f6}
\end{figure*}

\begin{deluxetable*}{ccccccccccccc}
\tablenum{1}
\tablecaption{Best-fitting Parameters for the log$_{10}$(\textrm{SSFR})$-$r$_{p}$ of the Star-forming Sequence in the Groups and Clusters \label{tab:ssfrfitting}}
\tablehead{
\colhead{}  & \multicolumn{5}{c}{0.2 $<$ $z$ $<$ 0.5}   & \colhead{} & \multicolumn{5}{c}{0.5 $<$ $z$ $<$ 0.8}  \\
\cline{2-6}\cline{8-12} \\
\colhead{$M_{*}$ Range}  & \multicolumn{2}{c}{Groups (610)$^{a}$}  & \colhead{}  & \multicolumn{2}{c}{Clusters (76)$^{a}$} & \colhead{}   & \multicolumn{2}{c}{Groups (875)$^{a}$}  & \colhead{} & \multicolumn{2}{c}{Clusters (61)$^{a}$}  \\
\cline{2-3}\cline{5-6}\cline{8-9}\cline{11-12} \\
\colhead{$log_{10}(M_{*}/M_{\odot})$} & \colhead{$\alpha$} & \colhead{$\beta$} & \colhead{}  & \colhead{$\alpha$}  & \colhead{$\beta$} & \colhead{}  & \colhead{$\alpha$} & \colhead{$\beta$} & \colhead{} & \colhead{$\alpha$}  & \colhead{$\beta$} \\
\cline{2-3}\cline{5-6}\cline{8-9}\cline{11-12}  
}
\startdata
9.7 - 10.1   & 0.035$\pm$0.016 & -9.605$\pm$0.024 & & 0.082$\pm$0.033 & -9.634$\pm$0.046 & & --- & --- & & --- & ---  \\
10.1 - 10.5 & 0.048$\pm$0.021 & -9.796$\pm$0.028 & & 0.103$\pm$0.024 & -9.887$\pm$0.034 & & 0.033$\pm$0.005 & -9.551$\pm$0.009 & & 0.092$\pm$0.047   & -9.692$\pm$0.080  \\
10.5 - 10.9 & 0.068$\pm$0.017 & -9.972$\pm$0.024 & & -0.114$\pm$0.054 & -9.771$\pm$0.035 & & 0.014$\pm$0.021 & -9.668$\pm$0.037 & & 0.093$\pm$0.018   & -9.806$\pm$0.020  \\
  $>$10.9   & 0.014$\pm$0.003 & -9.879$\pm$0.004 & & 0.068$\pm$0.067 & -9.915$\pm$0.039 & & 0.051$\pm$0.011 & -9.842$\pm$0.014 & & 0.015$\pm$0.033 & -9.858$\pm$0.012 \\
\enddata
\tablecomments{ $^{a}$This number denotes the number of groups or clusters used in the analysis of each subsample.}
\end{deluxetable*}

\begin{figure*}
\includegraphics[scale=1.4]{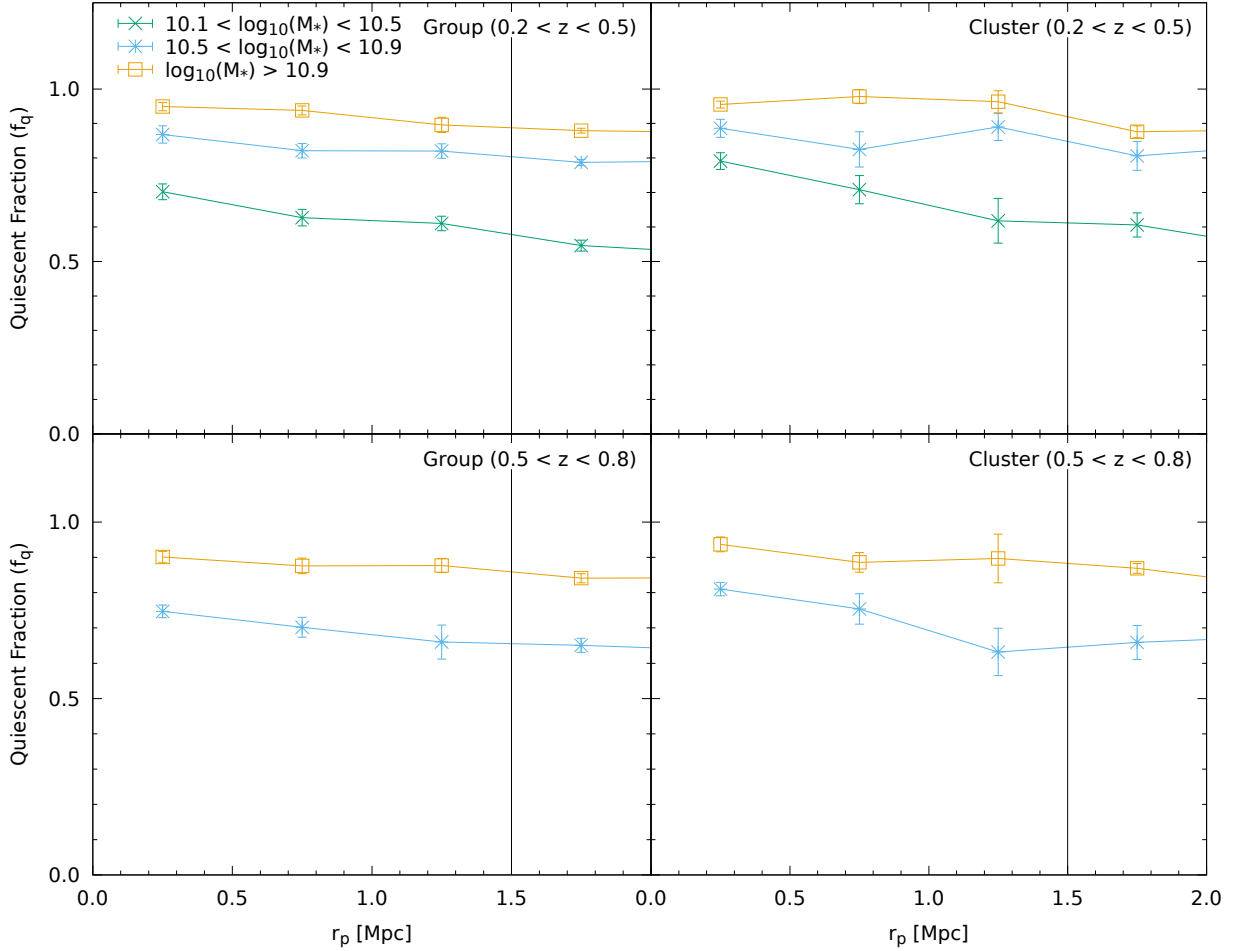}
\caption{Quiescent fraction $f_q$ is plotted as a function of $r_p$ in three or two different stellar mass ranges. The quiescent fraction, in general, slightly drops with the increasing group-centric radius. In the same environment, the slope of $f_q$ is steeper at lower redshift, while at the same redshift, the reduction of $f_q$ with the increasing $r_p$ is sharper in the clusters than in the groups. Moreover, the slope of the $f_q$s for less massive bins is steeper than the slope for more massive bins, implying that a stronger environmental effect acts on less massive galaxies.}
\label{f7}
\end{figure*}

\begin{deluxetable*}{ccccccccccccc}
\tablenum{2}
\tablecaption{Best-fitting Parameters for the $f_{q}$$-$r$_{p}$ in the Groups and Clusters \label{tab:fqfitting}}
\tablehead{
\colhead{}  & \multicolumn{5}{c}{0.2 $<$ $z$ $<$ 0.5}   & \colhead{} & \multicolumn{5}{c}{0.5 $<$ $z$ $<$ 0.8}  \\
\cline{2-6}\cline{8-12} \\
\colhead{$M_{*}$ Range}  & \multicolumn{2}{c}{Groups (610)$^{a}$}  & \colhead{}  & \multicolumn{2}{c}{Clusters (76)$^{a}$} & \colhead{}   & \multicolumn{2}{c}{Groups (875)$^{a}$}  & \colhead{} & \multicolumn{2}{c}{Clusters (61)$^{a}$}  \\
\cline{2-3}\cline{5-6}\cline{8-9}\cline{11-12} \\
\colhead{$log_{10}(M_{*}/M_{\odot})$} & \colhead{$\alpha$} & \colhead{$\beta$} & \colhead{}  & \colhead{$\alpha$}  & \colhead{$\beta$} & \colhead{}  & \colhead{$\alpha$} & \colhead{$\beta$} & \colhead{} & \colhead{$\alpha$}  & \colhead{$\beta$} \\
\cline{2-3}\cline{5-6}\cline{8-9}\cline{11-12}  
}
\startdata
10.1 - 10.5 & -0.098$\pm$0.013 & 0.720$\pm$0.017 & & -0.127$\pm$0.016 & 0.818$\pm$0.016 & & --- & --- & & --- & ---  \\
10.5 - 10.9 & -0.048$\pm$0.009 & 0.871$\pm$0.014 & & -0.037$\pm$0.031 & 0.894$\pm$0.003 & & -0.060$\pm$0.007 & 0.760$\pm$0.001 & & -0.114$\pm$0.023   & 0.834$\pm$0.017  \\
  $>$10.9   & -0.049$\pm$0.005 & 0.966$\pm$0.008 & & -0.046$\pm$0.025 & 0.974$\pm$0.023 & & -0.038$\pm$0.007 & 0.911$\pm$0.010 & & -0.041$\pm$0.010 & 0.939$\pm$0.014  \\
\enddata
\tablecomments{ $^{a}$This number denotes the number of groups or clusters used in the analysis of each subsample.}
\end{deluxetable*}

\begin{figure*}
\includegraphics[scale=1.4]{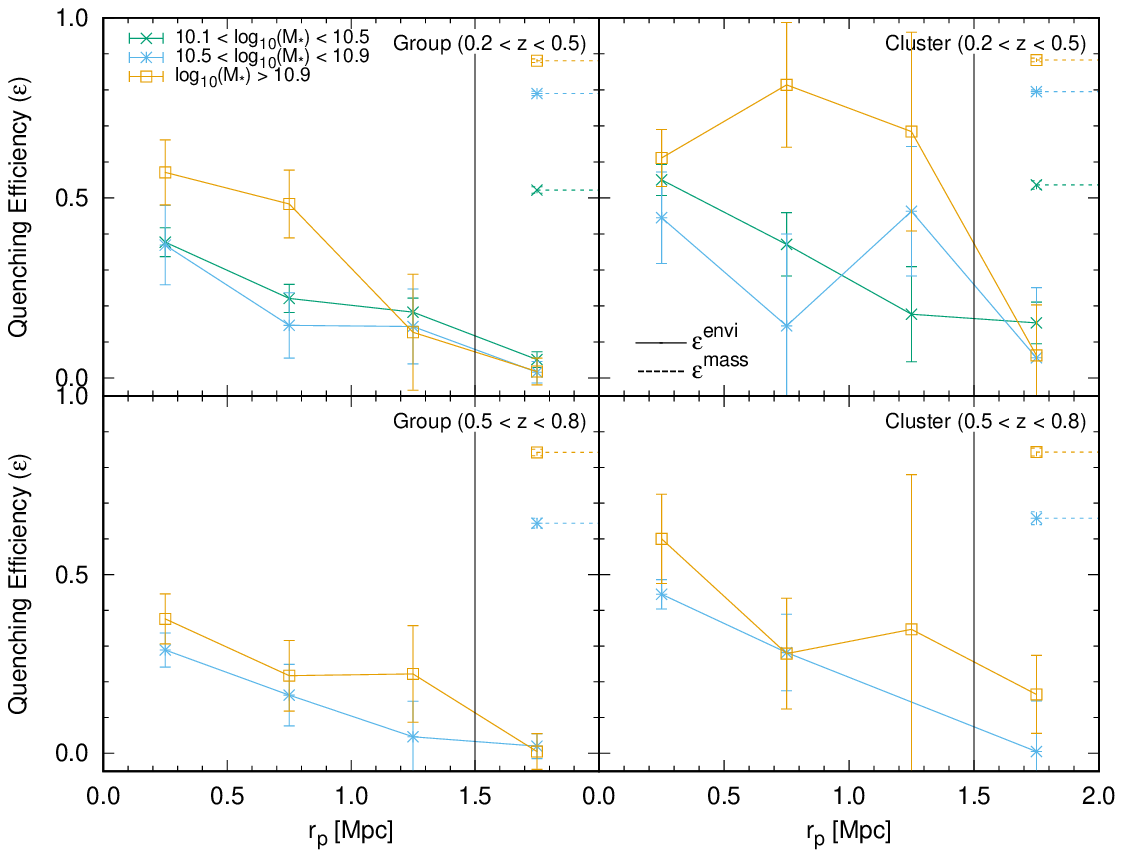}
\caption{Quenching efficiency $\varepsilon$ as a function of $r_p$. The solid lines show the environment-quenching efficiency, while the dashed lines trace the mass-quenching efficiency, equal to the field quiescent fraction by definition. The environment quenching increases with the decreasing radius. At all masses, the mass-quenching effect dominates the environmental quenching effect.}
\label{f8}
\end{figure*}

\subsection{Density Estimation}
In addition to the ``background subtraction" method, we also estimate the galaxy overdensity using the $n^{th}$-nearest-neighbor \citep[e.g.][]{coo07} approach to probe the color$-$density relation as well as to compare with results from the color as a function of the group-centric radius. The overdensity is computed from the surface density normalized by the median surface density for the redshift range where the surface density of a galaxy is derived from the area enclosed by the $sixth$ nearest neighbor on a projected plane from surrounding galaxies within a redshift slice equal to the photometric redshift uncertainty of the sample, i.e. the $z$ slice = $\pm$ $\sigma_{\Delta z/(1+z_s)}$. The photo-$z$ uncertainty of our galaxy sample is $\sigma_{\Delta z/(1+z_s)}$ $\sim$ 0.05 \citep{lin14}, and that is, the $z$ slice = $\pm$ 0.05. In addition, from simulations, it has been demonstrated that photometric redshift datasets are promising for detecting the color$-$density relation \citep{lai16}, and a similar conclusion is made for tests made out to redshift $z$ $\sim$ 2.5 \citep{lin16}. The density measure based on a photometric redshift dataset is therefore reliable and allows us to study the color$-$density relation in this work.  

\section{Results}
\subsection{Radial Dependence}
\subsubsection{Specific Star Formation Rate, SSFR}
We first investigate the radial dependence of the galaxy properties, i.e. study the galaxy properties as a function of a projected group-centric radius $r_p$ at fixed stellar mass, and their evolution in time. In Figure~\ref{f6}, we plot the median specific star formation rates (SSFRs) as a function of the projected group-centric radius $r_p$ in different stellar mass ranges for star-forming galaxies (solid lines) and all galaxies (dashed lines) in group (left) and cluster (right) environments and in the lower-$z$ bin $0.2 < z < 0.5$ (top) and in the higher-$z$ bin $0.5 < z < 0.8$ (bottom). Four stellar mass bins we considered: $9.7 < \textrm{log}_{10}(M_s/M_\odot) < 10.1$ (purple), $10.1 < \textrm{log}_{10}(M_s/M_\odot) < 10.5$ (green), $10.5 < \textrm{log}_{10}(M_s/M_\odot) < 10.9$ (blue), and $\textrm{log}_{10}(M_s/M_\odot) > 10.9$ (gold). The threshold dividing the ``star-forming'' and ``quiescent'' galaxies is SSFR = $10^{-10.1}~$yr$^{-1}$ in 0.2 $< z <$ 0.5 and $10^{-10.0}~$yr$^{-1}$ in 0.2 $< z <$ 0.5. Because of the redshift-dependent mass completeness limits for the star-forming and quiescent galaxies, the star-forming and all galaxy samples at low-$z$ are probed in four and three bins, respectively, while at high-$z$, there are only three bins for the star-forming galaxies and two bins for all galaxies. 

To quantitatively describe our results, the SSFRs of star-forming galaxies are fit with a linear relation log$_{10}$(SSFR/yr$^{-1}$) = $\alpha$ $\times$ ($r_p$/Mpc) + $\beta$ for $r_p$ $<$ 2.0 \rm{Mpc} and the best-fit parameters are listed in Table~\ref{tab:ssfrfitting}. At fixed mass, the SSFR deficit of group to field galaxies is $<$ 0.1 dex, consistent with our previous finding in \cite{lin14}. That is, the median SSFRs of star-forming galaxies in the group environment in both redshift bins are roughly independent of the $r_p$, suggesting that the environmental quenching effect is likely dominated by a fast mechanism. However, we note that the 0.1 dex reduction of the SSFR for low-mass group galaxies has a 2.5$\sigma$ confidence and is small, but not totally negligible. On the other hand, in the cluster environment, the SSFRs of the star-forming galaxies show strong scatter between the center and field, with a difference $<$ $\sim$ 0.2 dex in less massive galaxy bins and roughly no reduction in massive galaxies. The SSFR of the low-mass star-forming sequences in the cluster environment on average reveal a more apparent drop than those in the group environment, indicating that star-forming cluster galaxies are possibly quenched by a slow effect. Assuming that the SSFR of field galaxies is reduced by the amount of suppression seen in the cluster center for low-mass SF galaxies, i.e. 0.2 dex, the $f_q$ of field galaxies increases from $\sim$ 0.54 to 0.71, but it is lower than the center $f_q$ 0.79, implying that the amount of SSFR reduction cannot fully account for the excess in the quiescent fraction relative to the field, and a fast quenching process probably also operate in low-mass cluster galaxies. In addition, at fixed radius, the SSFRs of the star-forming galaxies strongly correlate with the stellar mass, suggesting that the primary factor in determining the SSFRs of star-forming galaxies is their stellar mass, not their location inside groups or clusters.

By contrast, at fixed mass, the SSFRs of all galaxies, including both the star-forming and quiescent galaxies, decrease significantly toward the center for less massive bins and roughly remain roughly constant for massive bins, implying that less massive galaxies suffer a stronger environmental effect than massive galaxies. The massive galaxies are likely old and dead before they are accreted on to dense environments, leading to them being less strongly affected by the environmental quenching and thus mainly dominated by the mass-quenching effect. In addition, at fixed radius, the median SSFR of all galaxies also drops significantly as the stellar mass increases more rapidly than for the star-forming galaxies. Learning from the median SSFRs of all and star-forming galaxies, the main environmental effect appears to move the less massive group or cluster galaxies out of the star-forming sequence to the quiescent population, leading to a suppression of the mean SSFR of all group or cluster galaxies. 

\subsubsection{Quiescent Fraction $f_q$}
Similar to Figure~\ref{f6}, the quiescent fractions ($f_q$s) are plotted in Figure~\ref{f7}. The $f_q$s are also fit with a linear relation $f_q$ = $\alpha$ $\times$ ($r_p$/Mpc) + $\beta$ for $r_p$ $<$ 2.0 Mpc,  and their best-fit parameter are listed in Table~\ref{tab:fqfitting}. There is one caveat to be noted: our comparisons are made at fixed physical cluster-/group-centric radius for different redshifts, but the physical sizes of clusters/groups change with time and also potentially with cluster/group richness. More discussion can be found in Section~\ref{dis}. From Figure~\ref{f7}, it is found that the quiescent fraction slightly decreases as the radius increases, by roughly a factor of 1 to 2 from the center to the field. On the basis of the PFOF identified group (or cluster) member galaxies, we split the groups (or clusters) into the red and blue groups (clusters), where the definition for the red groups is that the red fraction in a PFOF group is greater than 0.6, and for the blue groups it is lower than 0.4, to further study their quiescent fractions via background subtraction. We find that the $f_q$ of the red groups reveals a stronger drop than that of all groups, but conversely, the $f_q$ of the blue groups shows a reversed (or flatter) slope with respect to the slope of all groups. The nearly flat $f_q$s in Figure~\ref{f7} for less massive galaxies can be attributed to an average effect between the high $f_q$ from the red groups and the low  $f_q$ from the blue groups. The blue group selection naturally leads to a low $f_q$. It is also possible that the false identifications from linking line-of-sight blue field galaxies as a group makes the quiescent fraction low.  On the other hand, the high $f_q$ from the red groups reminds us that for groups (or clusters) that are identified using red-sequence methods, a steep $f_q$ slope is expected because of the bias that red-sequence groups or clusters need to have enough red galaxies, or equivalently, to have a high red fraction, to form the red sequence. 

In addition, the dependence of the quiescent fraction on the stellar mass is stronger than the dependence on the radius. The trend of increasing quiescent fraction with the decreasing radius appears to be weaker than the trend with the increasing stellar mass, suggesting that the stellar mass may control this ratio slightly more than the radius. The steepest radial changes of $f_q$ for all group mass and redshift bins are in the mass bin 10.1 $<$ $\textrm{log}_{10}(M_*/M_{\odot})$ $<$ 10.5, indicating that they are mostly affected by the group or cluster environment, consistent with the results found by \cite{li12}.  Additionally, when the quiescent fraction of the group and cluster galaxies is compared at the same redshift and fixed stellar mass, the cluster galaxies have a higher quiescent fraction than the group galaxies, and the quiescent fraction increases as the redshift decreases, exhibiting an apparent group downsizing effect \citep{li12}. Under the mass control, the group downsizing effect is still visible for less massive member galaxies, implying that these lower-mass member galaxies in clusters are at a more advanced evolutionary state than those in groups. On the basis of Table~\ref{tab:fqfitting}, it is seen that for less massive galaxies at low redshift, the best-fit slope of the quiescent fraction is -0.127 in clusters, steeper than the slope -0.098 in groups. Similarly, at high redshift, it is -0.114 in clusters, steeper than -0.06 in groups. A trend is seen that at the same redshift for less massive galaxies, the slope of the quiescent fraction appears to be slightly steeper in the cluster than in the group galaxies, implying that the environmental quenching effect is stronger in the cluster. A similar conclusion was made in \cite{li12}. It is also observed in groups and clusters that $f_q$s increases toward lower redshift at fixed mass and radius, consistent with the Butcher$-$Oemler effect \citep{but84}. The Butcher$-$Oemler effect appears to be weak in our sample, and the weak $f_q$ dependence on redshift is also observed in \cite{chiu16} using a 46 X-ray selected group sample from redshift 0.1 up to $\sim$ 1.0. 

\subsubsection{Quenching Efficiency $\varepsilon$}

Following \cite{lin14}, we also compute the environmental quenching efficiency, $\varepsilon^{\textrm{envi}}$ = ($f^{\textrm{group}}_{q} - f^{\textrm{field}}_{q}$)/(1 - $f^{\textrm{field}}_{q}$) \citep{pen10}, as a function of group-centric radius in different stellar mass bins to quantify the excess of quenching that is due to pure environmental effects in Figure~\ref{f8}. The $\varepsilon^{\rm{envi}}$s depends on the radius. The environment-quenching efficiency drops as the radius increases. The level of $\varepsilon^{\textrm{envi}}$ is lower in the higher-$z$ bin and for group galaxies, suggesting that the act of environment quenching operates more strongly in the local universe than at higher redshifts and in the cluster environment than in the group environment. In addition, the $\varepsilon^{\textrm{envi}}$ also shows a stellar mass dependence. The environment quenches star formation more efficiently for galaxies with higher stellar mass, but mass quenching still dominates environmental quenching at high mass. On the other hand,  the environmental quenching at the center becomes comparable to the mass quenching in less massive galaxies, although their environmental quenching efficiency is less effective than the high-mass quenching.

\begin{figure}
\includegraphics[scale=0.6]{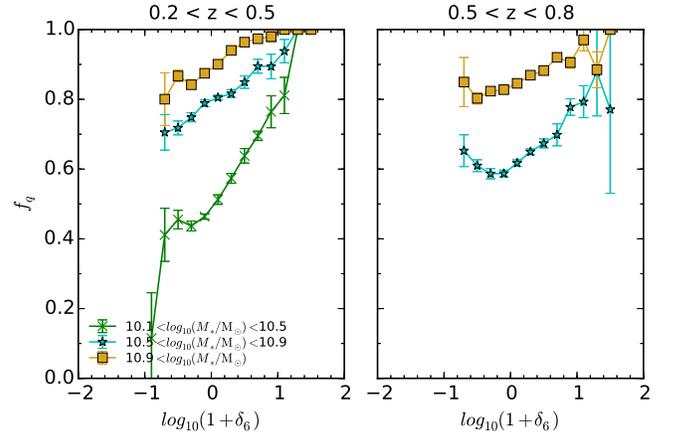}
\caption{The color$-$density relation. Similar to Figure \ref{f7}, but the quiescent fraction $f_q$s are plotted as a function of the overdensity $\textrm{log}_{10}(1+\delta_6)$. }
\label{f9}
\end{figure}

\subsection{Density Dependence}\label{dd}
\begin{figure*}
\includegraphics[scale=1.4]{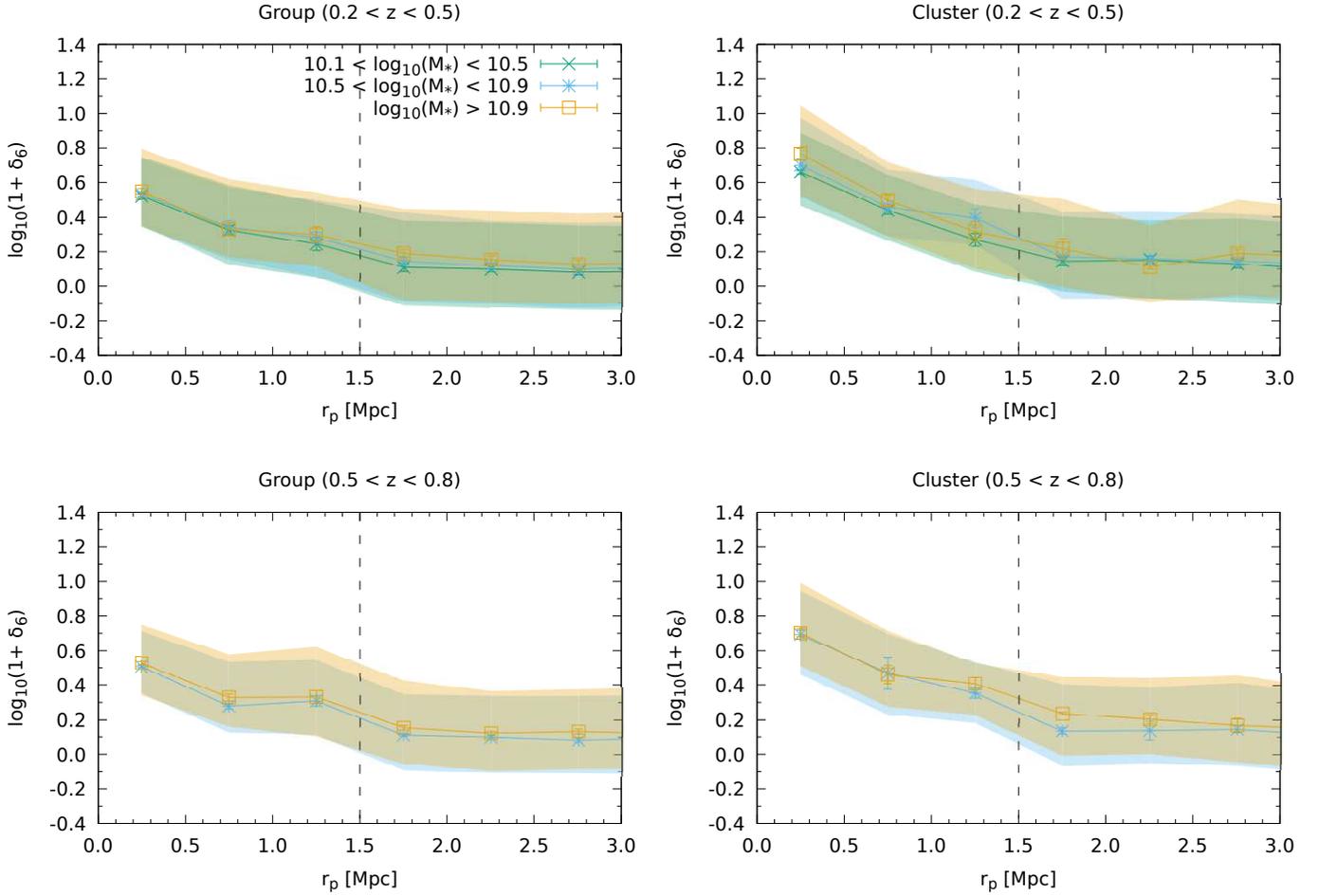}
\caption{ Overdensity $\textrm{log}_{10}(1+\delta_6)$ as a function of the group-centric radius $r_p$. The solid lines show the median densities for three different mass ranges at low-$z$ and two bins for high-$z$ while their corresponding color shaded regions indicate the density between 20th and 80th percentile. The median densities depend only on the radius, not on the stellar mass.}
\label{f10}
\end{figure*}

Previous studies have shown that the local galaxy density plays a role in transforming star-forming galaxies into passive galaxies \citep[e.g.][]{goto03}.  Since the galaxy density in groups or clusters drops with the increasing radius, it is expected that the effects of density and $r_p$ very likely interact with each other. By examining how these two effects directly correlate with each other in our sample, we may remove the quiescent fraction caused by the density effect and obtain the quiescent fraction that is purely due to the radial effect.  By adopting the $n$th-nearest-neighbor approach \citep{coo07}, with $n$ = 6 in this work,  we compute the quiescent fraction as a function of over-density shown as in Figure~\ref{f9}. It can be seen that $f_q$ increases with the overdensity, consistent with what we obtained previously for the radial dependence, where the $f_q$ decreases with increasing $r_p$, and the galaxy density drops with increasing radius. The direct way to probe the relation between the radial and density effect is to explore galaxy properties as a function of $M_*$, $r_p$, and $\textrm{log}_{10}(1+\delta_6)$. However, because of the insufficient sample size, we instead compute the median overdensity as a function of the group-centric radius, as shown in Figure~\ref{f10} via the background subtraction method. The solid lines give the median densities for different masses, while the color shaded regions denote the density between the 20th and 80th percentile. The density gradually decreases with increasing radius, and the core density is higher in clusters than in groups, as expected.  In addition, the galaxy density at fixed $r_p$ is approximately independent of $M_*$  in groups or clusters at the same redshift, implying that galaxies with different masses have similar radial distributions. 

\begin{figure*}
\includegraphics[scale=1.2]{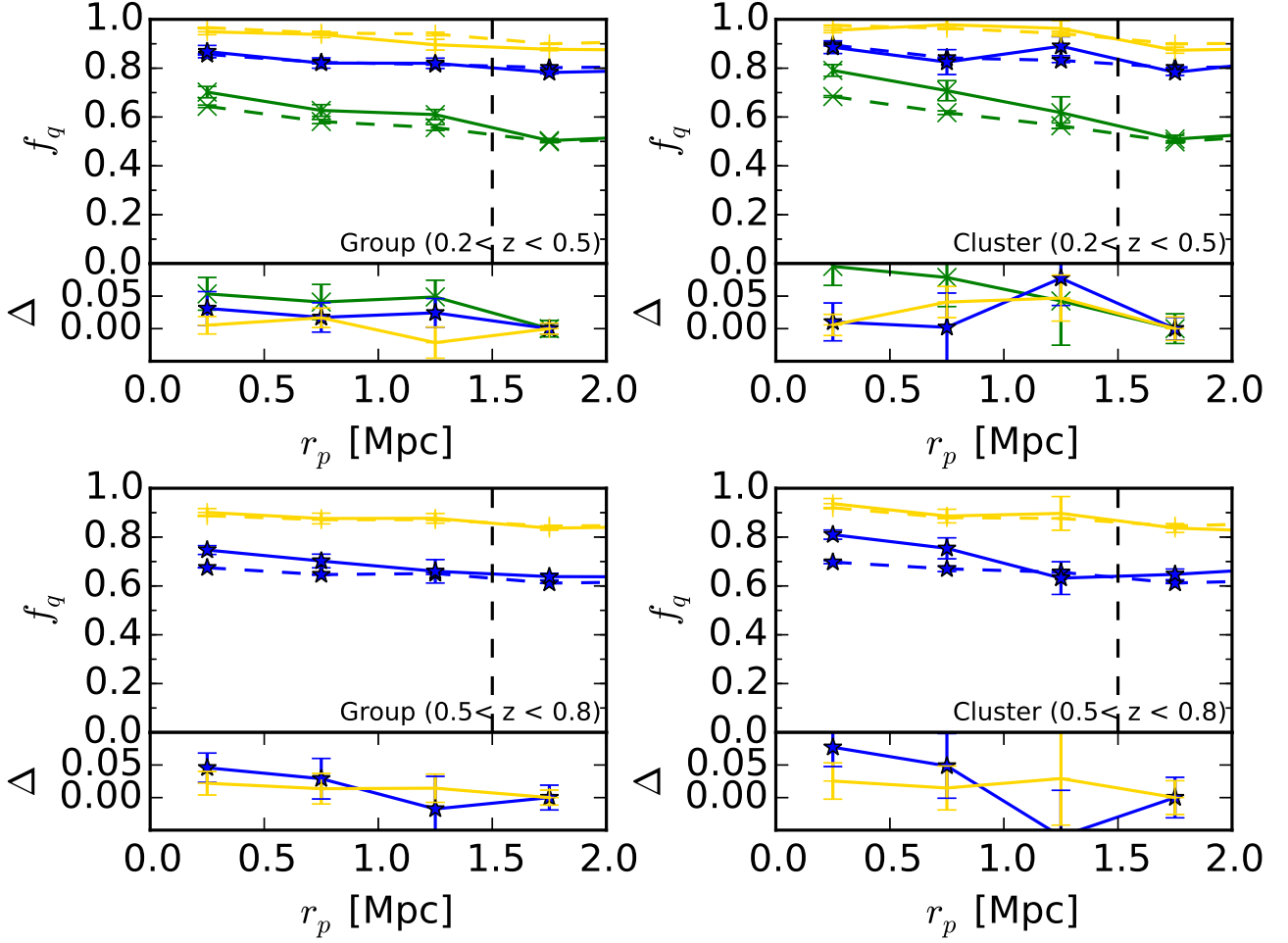}
\caption{Top: comparisons of the quiescent fraction between the results from Figure~\ref{f7} (solid lines) and from those combining the color-density relation in Figure~\ref{f9} and the radius$-$density relation in Figure~\ref{f10} (dashed lines). Three stellar mass ranges are plotted:  $10.1 < \textrm{log}_{10}(M_s/M_\odot) < 10.5$ (green), $10.5 < \textrm{log}_{10}(M_s/M_\odot) < 10.9$ (blue), and $\textrm{log}_{10}(M_s/M_\odot) > 10.9$ (gold). Bottom: The excess of the quiescent fraction ($\Delta$) from the radial (solid lines) and density (dashed lines) contribution relative to the fraction from the field.}
\label{f11}
\end{figure*}

Using the results in Figures \ref{f9} and \ref{f10}, we can convert the density dependence to the $r_p$ dependence into compare their relative roles in the quenching effect between the density effect and the radial effect. The results are shown at the top of in each panel of Figure~\ref{f11}. The solid lines are results from Figure~\ref{f7}, which contains contributions from the radial effect, while the dashed lines represent the quiescent fraction contributed purely by the density effect. At the bottom, we plot the excess of the $f_q$ ($\Delta$) from the radial (solid lines) relative to the density (dashed lines) contribution. For high-mass galaxies, the quiescent fractions can be solely accounted for the density effect, and this shows that the density effect appears to be the main quenching source for high-mass galaxies, suggesting that the quenching mechanism related to the density effect, i.e. merger or galaxy harassment, is likely the dominant quenching process. Moreover, less massive galaxies on average have a higher $f_q$ in the center of the cluster than the galaxies in the outskirts after removing the $f_q$ contribution from the density effect. In other words, the $\Delta$ in clusters has a radial gradient, implying that the location of the less massive galaxies in their parent cluster halo has an additional effect on their star formation quenching. Likewise, a weaker trend is also displayed in low-mass group galaxies. A similar result was also found by \cite{li12}: for lower mass galaxies inside groups or clusters, not only the density effect can contribute galaxies to the red population, but their locations also have an effect on turning galaxies red. The quenching mechanism related to the radial effect appears to act more efficiently in lower mass galaxies than in higher mass galaxies. Moreover, \cite{peng15} reported that for local galaxies with a stellar mass of $M_{\rm{star}}$ $<$ $10^{10}$ $M_{\odot}$, the primary quenching mechanism is strangulation. Hence, it is expected that the ram-pressure stripping as the dominant effect is less likely, and the starvation should be the leading quenching mechanism for these less massive galaxies. 

\begin{figure*}
\includegraphics[scale=1.4]{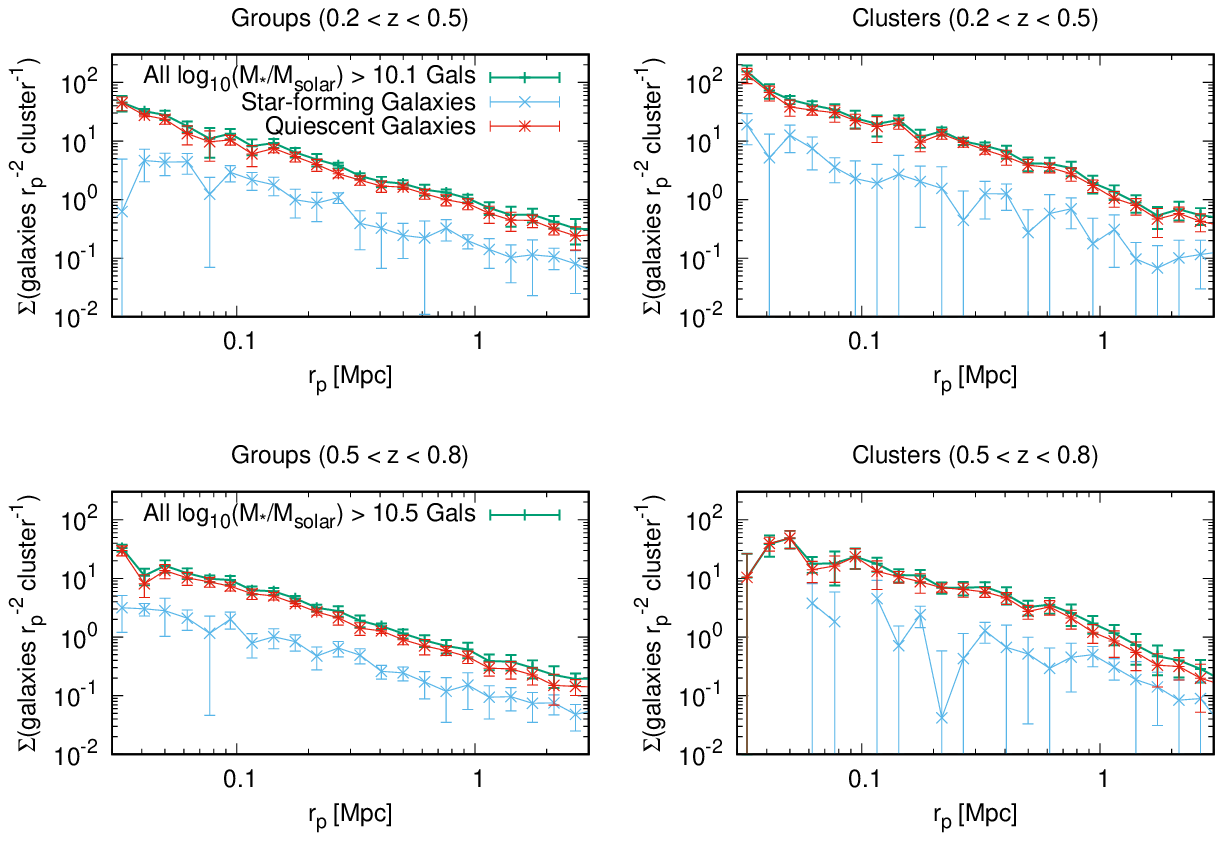}
\caption{Radial galaxy surface density profiles, $\Sigma(r)$, of all galaxies (green) with $log_{10}(M_*/M_{\odot})$ $>$ 10.1 (low-$z$) and 10.5 (high-$z$), the star-forming galaxies (blue), and the quiescent galaxies (red). The surface density of star-forming group (or cluster) galaxies falls steadily with the increasing radius out to ∼3 \rm{Mpc}, with no evidence of flattening off inside 0.5 \rm{Mpc}.}
\label{f12}
\end{figure*}

\section{Discussion}\label{dis}
It is known that the infall time of a satellite is linked to its radial distance from the core through dynamical friction \citep{gao04}. The build-up of quiescent galaxies through an environmental quenching process translates over time into the radial gradient in the quiescent fraction. Owing to the difference in quenching timescale for different processes, it is expected that a fast quenching mechanism can leave a different feature on $f_q(r)$ from a slow mechanism. It is found in simulations that models with short quenching timescales yield steeper cluster-centric gradients in disk colors and Balmer line indices than those with long quenching timescales \citep{tar14}. For a fixed period of time, a fast mechanism is therefore anticipated to produce more quiescent galaxies and consequently also leads to a sharper gradient in the $f_q(r)$ than a slow process. The longer time delay in a slow quenching process than in a fast process also implies that under such an environment, galaxies are expected to evolve passively for a longer time when accreted. That is, the profile of the quiescent fraction that is due to a slow quenching process will be significantly affected by passive evolution, likely resulting in a flatter slope than the fast process. With information from the $f_q$ alone, their absolute slopes are unknown, although the quenching effects due to two mechanisms may lead to different slopes of $f_q$. The two effects can be separated when we additionally consider the distribution of star-forming galaxies that is caused by these two different mechanisms. It is expected that the slow quenching process will deplete the SFR of the entire population and induce a radial gradient in the median SSFR of star-forming galaxies, while the fast process will truncate the star formation in a relatively short period of time, without altering the distribution of the star-forming population, and this will lead to a radially constant SSFR.

From Figure \ref{f6} and \ref{f7}, our results show an SSFR decline of $\sim$ 0.1 dex for low-mass galaxies in groups and suggest that inside the group environment, the quenching effect more likely fits the scenario of the fast process, and mainly acts to raise the fraction of the quiescent population, rather than to decrease the SFR of the entire population since the SSFRs of the star-forming galaxies in groups is independent of the group-centric radius. This finding is in agreement with the previous works \citep{vul10,muz12,wet12,koy13,koy14}. The quenching scenario appears to fit the delayed-then-rapid process proposed by \cite{wet12} that satellites remain actively star-forming on a timescale of roughly Gyr after the first accretion, unaffected by their host halo, before quenching starts, and once quenching has started, SFR fades with a relatively short timescale roughly several tenths of Gyr \citep{wet13,muz14}. However, our results do not entirely agree on the unchanged SSFR condition during the delayed process since the SSFR of star-forming group galaxies decreases slightly toward the core, and thus appears to not completely satisfy the claim. Of the possible quenching mechanisms, such as ram-pressure stripping \citep{gun72}, strangulation \citep{lar80}, galaxy harassment \citep{moo96}, and galaxy-galaxy mergers \citep{mih94}, the most likely mechanisms acting in the group environment are mergers and starvation. For the case of mergers, it is mainly due to its relatively short quenching timescale $<$1 Gyr \citep{lot10} and higher merger rate in denser environments \citep{lin10}, peaked at the group environment \citep{jian12}, and for the case of starvation, it is still effective in low-mass groups, although it is with a slightly longer quenching timescale $\geq$ 1.5 Gyr \citep{mcc08}, in contrast to ram-pressure stripping, which is less influential in groups \citep{fuj01}, and galaxy harassment, which occurs preferentially beyond the virial radius \citep{tre03,mor07}.

On the other hand, in the cluster environment, an SSFR depletion of star-forming galaxies of $\sim$0.2 dex is seen in low-mass star-forming cluster galaxies and roughly no reduction in massive galaxies compared to the field galaxies. The result is consistent with previous works \citep{vul10,hai13,alb14}. For massive star-forming cluster galaxies, no SSFR reduction implies that the main environmental effect is likely a fast process. On the other hand, for less massive star-forming galaxies, a 0.2 dex SSFR reduction can be attributed to a slow quenching effect. Our results also show that if the reduction in low-mass galaxies is due to a slow quenching effect, the slow quenching process acts primarily in clusters. The high-velocity dispersion in clusters makes the merger mechanism inefficient, but the high-density cluster environment increases the harassment frequency, making the harassment a possible quenching mechanism. In addition, the starvation as a slow quenching process is active in clusters and is also a potential quenching mechanism. 

In addition, from the results found in Section~\ref{dd}, we learn that more massive galaxies are more likely environmentally quenched by the density effect, while less massive galaxies are increasingly affected by the radial effect. Therefore, we conclude that the dominant environmental quenching effect for more massive galaxies is likely mergers in the group environment and is the harassment in clusters. The less massive group and/or cluster galaxies are primarily quenched by starvation. We also estimate the environmental variations of the fraction of starburst galaxies for $M_*$ $>$ 10$^{10.1}$ $M_{\odot}$, where starbursts are defined as galaxies with the SSFR $>$ 2 $\times$ SSFR of the main sequence \citep{elb11}. It is found that at high redshift, the fraction is $\sim$14$\%$ in groups and $\sim$12$\%$ in clusters, while at low redshift, it is $\sim$12$\%$  in star-forming group galaxies and $\sim$9$\%$ in cluster galaxies. That is, star-forming galaxies in groups have a higher starburst fraction than those in clusters, favoring a merger-induced starburst scenario because the velocity dispersion of galaxies in clusters could be too high for mergers.

Moreover, we also study the radial galaxy surface density to further understand the possible quenching mechanism in groups or clusters, as shown in Figure~\ref{f12}.  The green, blue, and red line denotes the surface density for all galaxies with the log $M/h^{-1}M_{\odot}$ larger than 10.1 at low-$z$ and 10.5 at high redshift for star-forming galaxies and for quiescent galaxies, respectively. Figure~\ref{f12} shows that the surface density of star-forming group and cluster galaxies increases steadily with decreasing radius from $\sim$ 3 Mpc to the center, without apparent flattening inside 0.5 Mpc. It is expected that models in which star formation is instantaneously quenched when galaxies are accreted into groups (or clusters) will produce flat radial profiles within 2$r_{500}$ \citep{hai15}. The fast quenching process, the ram-pressure stripping, thus appears to be less likely as the main environmental quenching mechanism inside clusters, and the most probable quenching mechanisms in groups and clusters seems to be the relatively slow and/or slow quenching process, i.e. mergers and/or starvation, consistent with our previous conclusions. 

Some issues related to the profile of the quiescent fraction are still worth some discussion. First, since the virial radius cannot be robustly derived in our sample, the cluster-centric radius can not be normalized by the virial radius, and the quiescent fraction at the boundary may be slightly smeared. The mass of our sample log$_{10}(M_h/\textrm{M}_{\odot})$ is 13.2-13.8 in the low-$z$ bin and is 13.4-14.0 in the high-$z$ bin. This would translate roughly into a factor of 1.6 in the radial range. If there is a sharp truncation in the outskirts, this might smear out the feature. On the other hand, a smooth trend with radius is probably not affected that much. The other issue is how does the time delay between ram-pressure stripping the gas and the stellar population changing color compares to the orbital crossing time. It is argued that when the time delay is comparable to the orbital crossing time, the galaxy would still be blue near the center when the gas is stripped, so one would not expect a strong color gradient. In contrast, if the galaxy immediately becomes red when its gas is being stripped near the center, it would remain red when moving out, which also lowers the color gradient. It is concluded that either way will dilute the radial dependence of the quiescent fraction. Because of the limitation of the current sample, our sample is not deep enough to let us probe the low-mass part at higher redshift, and the sample size is also too small to contain enough massive clusters to give better statistics. It is expected that the full PS1 MD data, which cover $\sim$ 70 deg$^2$, or HSC data can provide a large enough sample to help us to study these regimes.


\section{Summary}

We used a catalog of 1600 galaxy groups produced by the PFOF algorithm in two Pan-STARRS1 medium-deep fields to study the radial dependence of the group galaxy properties, i.e. the SSFR and quiescent fraction, for galaxies with stellar mass log$_{10}(M_*/M_{\odot}) \geqslant 10.1$ over the redshift range 0.2  $<$ $z$ $<$ 0.8. Adopting a stacking technique plus background contamination removal, we extended our previous study in \cite{lin14} to explore the radial dependence of the SSFR and quiescent fraction, $f_q$ of group galaxies in more detail. Since the density and radius effect are expected to interact with each other, we estimated the separate contributions from the density effect and from the pure radial effect on $f_q$ in order to understand the dominant quenching mechanism. Our results are summarized as follows:
\begin{enumerate}
\item The median SSFR decreases from the field toward the center, and the drop is more apparent for less massive galaxies than for more massive galaxies, for cluster galaxies more than for group galaxies, and for galaxies at lower $z$ more than for galaxies at higher $z$.  The relative difference of the median SSFR between group galaxies and field galaxies is $\sim$ 0.1 dex for the less massive bin, implying that fast quenching is likely the dominant environmental effect. On the other hand, the reduction in the SSFR for cluster galaxies is $\sim$ 0.2 dex for the less massive bin,  suggesting that the fast and slow quenching effect are likely acting comparably. 
\item The quiescent fraction, in general, falls with increasing radius, indicating that the environmental effect is stronger at the center than at the boundary. The $f_q$ slope is steeper for less massive bins than for more massive bins, showing that the less massive group or cluster galaxies suffer a stronger environmental effect. The flat $f_q$s for more massive group or cluster galaxies are likely due to the galaxies being old and dead and not significantly affected by the environmental effect. In addition, the dominated quenching effect appears to be mass quenching for more massive bins. 
\item Because of the small depletion of SSFR in group galaxies in contrast to the field galaxies, the main quenching process in groups is likely a fast mechanism, and the result favors galaxy mergers as being the main quenching mechanism. On the other hand, the depletion in cluster galaxies is $\sim$ 0.2 dex, implying that slow quenching processes might play a role. When the reduction is attributed to the slow quenching effect, the slow quenching processes appear to act dominantly in clusters. Strangulation is a plausible mechanism.
\item The environment-quenching efficiency $\epsilon$ is higher for more massive galaxies than for less massive galaxies. However, high-mass galaxies are dominated by the mass-quenching effect rather than the environment, and low-mass galaxies are quenched roughly comparably by the environment and the mass.  In addition, at fixed mass, galaxies in clusters show a higher  $\epsilon$ than those in groups, and low-$z$ group (or cluster) galaxies have a higher $\epsilon$ than those at high-$z$.

\item  In the more massive group or cluster galaxies, the environmental effect can be primarily accounted for a similar way by the density and radial effect. On the other hand, in less massive galaxies, the radial environmental effect dominates over the density effect in groups and clusters.    
\end{enumerate}

Our sample covers redshifts up to 0.8, and to truly understand galaxy evolution, we need to extend the redshift to higher redshift, $z$ $>$ 1, to contain a wide redshift range. In addition, we also need a sufficiently large sample to have better statistics and clearly separate the effects of these parameters. Moreover, both theoretical predictions and observations of the very nearby universe both suggest that low-mass galaxies (log$_{10}[M_*/M_{\odot}] < 9.5$) are likely to remain forming stars unless they are affected by their local environment. The low-mass galaxies in groups or clusters are very likely to have different dominant quenching mechanisms to the high-mass galaxies. Because of the mass completeness limit of our current sample, we can only study galaxies with mass $>$ $10^{10} M_{\odot}$, and this restricts our exploration at low-mass galaxies to invest the premise observationally in more detail. The recent HSC survey appears to satisfy our needs and will allow us to make great strides toward this goal.

\vspace{3mm}

\emph{Acknowledgments}-

We thank Howard Yee and Surhud More for helpful discussions. The work is supported in part by the National Science Council of Taiwan under the grant MOST 103-2112-M-001-031-MY3, NSC101-2112-M-001-011-MY2, and NSC102-2119-M-008-001. S.C. acknowlges the support of STFC [ST/F00075X/1]. The Pan-STARRS1 Surveys (PS1) have been made possible through contributions of the Institute for Astronomy, the University of Hawaii, the Pan-STARRS Project Office, the Max Planck Society and its participating institutes, the Max Planck Institute for Astronomy, Heidelberg and the Max Planck Institute for Extraterrestrial Physics, Garching, The Johns Hopkins University, Durham University, the University of Edinburgh, Queen's University Belfast, the Harvard-Smithsonian Center for Astrophysics, the Las Cumbres Observatory Global Telescope Network Incorporated, the National Central University of Taiwan, the Space Telescope Science Institute, the National Aeronautics and Space Administration under Grant No. NNX08AR22G issued through the Planetary Science Division of the NASA Science Mission Directorate, the National Science Foundation under Grant No. AST-1238877, and the University of Maryland, and Eotvos Lorand University (ELTE). PS1 images and catalogs will be made available through a Pan-STARRS PS1 data release by STScI.

\end{document}